\documentclass[prl,aps,nofootinbib,superscriptaddress,showpacs,floatfix,preprintnumbers,twocolumn]{revtex4}
\usepackage{hyperref,amssymb,amsmath,graphicx,xcolor,slashed}

\DeclareMathOperator\Tr{Tr}

\begin{document}

\title{Valence-Quark Distribution of the Kaon and Pion from Lattice QCD}

\author{Huey-Wen Lin}
\email{hwlin@pa.msu.edu}
\affiliation{Department of Physics and Astronomy, Michigan State University, East Lansing, MI 48824}
\affiliation{Department of Computational Mathematics,
  Science and Engineering, Michigan State University, East Lansing, MI 48824}
  
\author{Jiunn-Wei Chen}
\email{jwc@phys.ntu.edu.tw}
\affiliation{Department of Physics, Center for Theoretical Physics, and Leung Center for Cosmology and Particle Astrophysics, National Taiwan University, Taipei, Taiwan 106}

\author{Zhouyou Fan}
\affiliation{Department of Physics and Astronomy, Michigan State University, East Lansing, MI 48824}
  
\author{Jian-Hui Zhang}
\email{zhangjianhui@bnu.edu.cn}
\affiliation{Center of Advanced Quantum Studies, Department of Physics,
Beijing Normal University, Beijing 100875, China}

\author{Rui Zhang}
\affiliation{Department of Physics and Astronomy, Michigan State University, East Lansing, MI 48824}
\affiliation{Department of Computational Mathematics,
  Science and Engineering, Michigan State University, East Lansing, MI 48824}

\preprint{MSUHEP-20-006}



\begin{abstract}
We present the first lattice-QCD calculation of the kaon valence-quark distribution functions using the large-momentum
effective theory (LaMET) approach. 
The calculation is performed with multiple pion masses with the lightest one around 220~MeV, 2 lattice spacings $a=0.06$ and 0.12~fm, $(M_\pi)_\text{min} L \approx 5.5$, and high statistics ranging from 11,600 to 61,312 measurements. 
We also calculate the valence-quark distribution of pion and find it to be consistent with the FNAL E615 experimental results, and our ratio of the $u$ quark PDF in the kaon to that in the pion  agrees with the CERN NA3 experiment. We also make predictions of the strange-quark distribution of the kaon. 
\end{abstract}

\maketitle

\section{Introduction}\label{sec:intro}
Light pseudoscalar mesons play a fundamental role in QCD as they are the Nambu-Goldstone bosons associated with dynamical chiral symmetry breaking (DCSB). While studies of pion and kaon structure both reveal physics of DCSB, a comparison between them helps to reveal the relative impact of DCSB versus the explicit breaking of chiral symmetry by the quark masses.  
An important quantity characterizing the structure of the pion and kaon is their parton distribution functions (PDFs). They can be measured by scattering a secondary pion ($\pi$) or kaon ($K$) beam over target nuclei ($A$), inducing the Drell-Yan process, $\pi (K)A \to X \mu^+ \mu^-$ 
~\cite{Badier:1983mj,Betev:1985pf,Falciano:1986wk,Guanziroli:1987rp,Conway:1989fs}. 
With a combined analysis of $\pi^{\pm}A$ Drell-Yan on the same nuclear target, the valence and sea distributions can be separated~\cite{Badier:1983mj}, provided that the nuclear PDF is known. Currently, the nuclear PDFs are approximated by a combination of proton and neutron PDFs. The valence-quark PDF of the pion for momentum fraction $x \gtrsim 0.2$ has been determined reasonably well~\cite{Badier:1983mj,Betev:1985pf,Conway:1989fs,Wijesooriya:2005ir,Aicher:2010cb}, subject to the systematic uncertainty in the PDF parametrization.

Combining $K^-A$ and $\pi^-A$ Drell-Yan data, the kaon valence PDF can be measured through the ratio~\cite{Badier:1980jq} $\bar{u}_v^{K^-}(x)/[\bar{u}_v^{\pi^-}(x)C(x)]$ 
where $\bar{u}_v^{K^-(\pi^-)}$ denotes the valence anti-up distribution in the $K^{-}$($\pi^{-}$). The function $C(x)$ encodes the corrections needed due to the nuclear modification of the target PDFs, the omission of meson sea-quark distributions and the ignorance of the ratio $s_v^{K^-}(x)/\bar{u}_v^{K^-}(x)$. In principle, the first two can be addressed by new experiments. For example, the valence and sea PDFs for the pion and kaon at $x > 0.2$ can be separated in the  $\pi^{\pm}$ and $K^{\pm}$ Drell-Yan experiments proposed by the COMPASS++/AMBER collaboration using the CERN M2 beamline~\cite{Denisov:2018unj}. Numerically, the biggest uncertainty in $C(x)$ is due to ignorance of the $s_v^{K^-}(x)/\bar{u}_v^{K^-}(x)$ ratio, and a reliable theoretical determination of this ratio, e.g. by lattice QCD, would greatly reduce the uncertainty in $\bar{u}_v^{K^-}(x)/\bar{u}_v^{\pi^-}(x)$.

Another experiment that could measure the pion and kaon PDFs is tagged deep inelastic scattering (TDIS), such as $ep\to e' (n\text{ or }Y)X$.
By tagging a neutron ($n$) or hyperon ($Y$) with specific kinematics in the final state of an $ep$ scattering, one can select events of the Sullivan process~\cite{Sullivan:1971kd},
where an electron scatters off an intermediate $t$-channel pion or kaon. 
Experimentally, the tagged-neutron DIS experiment was pioneered by HERA, covering $x < 0.01$~\cite{Aaron:2010ab}.
Approved experiments at JLab aim to determine $\bar{u}_v^{\pi^-}$ for $x >0.45$ with better than $1.1\%$ statistical and $6.5\%$ systematic uncertainty~\cite{Keppel:2015} and to determine $\bar{u}_v^{K^-}$ in the same range with $3\%$ statistical and $6.5\%$ systematic uncertainty~\cite{Keppel:2015B}. The combined result will determine the ratio with $3\%$ statistical and $5\%$ systematic uncertainty~\cite{Keppel:2015B}. 
At the future Electron-Ion Collider, TDIS experiments can cover from $x=10^{-3}$ with $Q^2=1\text{ GeV}^2$ to $x=1$ with $Q^2=1000\text{ GeV}^2$. The statistical uncertainty of the ratio $\bar{u}_v^{K^-}(x)/\bar{u}_v^{\pi^-}(x)$ can be reached to 1--$3\%$ level for $x \in [0.2,0.9]$ with about $5\%$ systematic uncertainty~\cite{Aguilar:2019teb}.   

Given the great experimental interest and effort to probe the pion and kaon PDFs, it is timely that these quantities have recently become calculable in lattice QCD, thanks to the development of large-momentum effective theory (LaMET)~\cite{Ji:2013dva,Ji:2014gla}. LaMET provides a general framework to extract lightcone correlations, such as the PDFs of hadrons, from equal-time Euclidean correlations calculable on the lattice. The latter can be computed at a moderately large hadron momentum, and then converted to the former through factorization formulas accurate up to power corrections that are suppressed by the hadron momentum.

Since its proposal, LaMET has been applied to computing various nucleon PDFs~\cite{Lin:2014zya,Chen:2016utp,Lin:2017ani,Alexandrou:2015rja,Alexandrou:2016jqi,Alexandrou:2017huk,Chen:2017mzz,Liu:2018hxv,Lin:2018qky,Lin:2019ocg}, the pion PDF and GPDs~\cite{Chen:2018fwa,Chen:2019lcm}, as well as the meson distribution amplitudes~\cite{Zhang:2017bzy,Chen:2017gck}, yielding encouraging results. 
In particular, the state-of-the-art calculation of the unpolarized and polarized isovector quark PDF of the nucleon~\cite{Chen:2018xof,Lin:2018qky} agrees with the global PDF fits~\cite{Dulat:2015mca,Ball:2017nwa,Accardi:2016qay,Nocera:2014gqa,Ethier:2017zbq} within errors. There have also been ongoing efforts to achieve full control of lattice systematics, including an analysis of finite-volume systematics~\cite{Lin:2019ocg} and exploration of machine-learning application~\cite{Zhang:2019qiq} that have been carried out recently. 
In parallel with the progress using LaMET, other proposals to calculate the PDFs in lattice QCD have also been formulated and applied to various parton quantities~\cite{Ma:2014jla,Ma:2017pxb,Radyushkin:2017cyf,Liu:1993cv,Liang:2017mye,Detmold:2005gg,Braun:2007wv,Bali:2017gfr,Chambers:2017dov}. Of course, each of them is subject to its own systematics.

In this paper, we carry out the first lattice-QCD calculation of the valence-quark distribution of the kaon using LaMET. Our calculation is done using clover valence fermions on an ensemble of gauge configurations with $N_f=2+1+1$ (degenerate up/down, strange and charm) flavors of highly improved staggered quarks (HISQ)~\cite{Follana:2006rc}, generated by the MILC Collaboration~\cite{Bazavov:2012xda} with two lattice spacings $a = 0.06$ and $0.12$~fm and three pion masses, approximately 690, 310 and 220~MeV.  To facilitate comparison with experimental results and other calculations, we also compute the valence-quark distribution of the pion using the same lattice setup.

\section{Kaon  and Pion PDFs from Lattice Calculation using LaMET}  \label{sec:lamet}
To see how the quark PDF in the kaon (or similarly for the pion) can be obtained within LaMET, we begin with the following operator definition
\begin{align}\label{pdfandqpdf}
&q^K(x) = \int\frac{d\lambda}{4\pi} e^{-i x \lambda n\cdot P}h_{\slashed n}(\lambda n) \\
& = \int\frac{d\lambda}{4\pi} e^{-i x \lambda n\cdot P} \langle K(P)\left|\bar\psi_q(\lambda n)\slashed n W(\lambda n,0)\psi_q(0)\right|K(P)\rangle,\nonumber
\end{align}
where $|K(P)\rangle$ denotes a kaon state with momentum $P^\mu=(P_t,0,0,P_z)$, $\psi_q, \bar\psi_q$ are the quark fields of flavor $q$, $n^\mu$ is a unit direction vector and $W\left(\zeta n, \eta n \right) =
  \exp [ig \int_\eta^\zeta d\rho\,n \cdot A(\rho n)]$ 
is the gauge link inserted to ensure gauge invariance. For later convenience, we have used a subscript $\slashed n$ on $h$ to denote the Dirac structure sandwiched between the quark fields. If we choose lightlike $n=n_{+}=(1,0,0,-1)/\sqrt{2}$, then Eq.~\ref{pdfandqpdf} defines the usual quark PDF with $x$ denoting the fraction of kaon momentum carried by the quark $q$. 
The support of $x$ is $[-1,1]$ with the negative $x$ part corresponding to the antiquark distribution: ${\bar q}^K(x)=-q^K(-x)$ for $x > 0$. One can define the 
valence-quark distribution for the positive range as $q_{v}^K(x)=q^K(x)-{\bar q}^K(x)$, which satisfies $\int_0^1 dx\, q_{v}^K(x)=1$ for a quark of the appropriate flavor.

On the other hand, if we choose spacelike $n=\tilde n=(0,0,0,-1)$, then Eq.~\ref{pdfandqpdf} becomes a Euclidean correlator known as quasi-PDF, which can be calculated in lattice QCD. For a given momentum $P_z\gg \Lambda_\text{QCD}$, the quasi-PDF has the same infrared physics as the PDF, so the two quantities can be connected via a factorization formula. Such a factorization can be done with either bare or renormalized correlators. In the present calculation we will follow the latter, since it facilitates the conversion from lattice results to results in the continuum.

On the lattice, we first calculate the quasi-PDF matrix element in coordinate space, and then renormalize it nonperturbatively in the regularization-independent momentum-subtraction (RI/MOM) scheme~\cite{Martinelli:1994ty}. To avoid potential mixing with scalar operators, we replace the Dirac structure $\slashed{\tilde n}$ with $\slashed{\tilde n_t}$, where $\tilde n_t=(1,0,0,0)$~\cite{Constantinou:2017sej,Chen:2017mie}. 
The RI/MOM 
renormalization factor $Z$ can be determined by demanding that it cancels all the loop contributions for the matrix element in an off-shell external quark state at a specific momentum, as was done in~\cite{Stewart:2017tvs,Chen:2017mzz}. 
After renormalization and taking the continuum limit, we can Fourier transform the renormalized matrix element to momentum space using Eq.~(\ref{pdfandqpdf}) and convert it to the lightcone PDF in $\overline{\text{MS}}$ scheme via the factorization
\begin{align}
\label{matching}
q^K(x,\tilde n,\tilde \mu)&=\int\frac{d y}{|y|}C\left(\frac{x}{y},\frac{\tilde \mu}{\mu},\frac{\mu}{y P_z}\right)
\tilde{q}^K(y, n_+,\mu)\nonumber \\
&+\mathcal O\left( \frac{m_K^2}{P_z^2},\frac{\Lambda_\text{QCD}^2}{x^2 P_z^2}\right),
\end{align}
where $C$ is a perturbative matching kernel 
that has been used in our previous works~\cite{Chen:2018xof,Lin:2018qky,Chen:2018fwa,Chen:2019lcm}.

\begin{table*}[tbp]
\begin{center}
\begin{ruledtabular}
\begin{tabular}{l|cccccccccc}
Ensemble ID & $a$ (fm) & $N_s^3 \times N_t$& $M_\pi^\text{val}$ (MeV) & $M_{\eta_s}^\text{val}$ (MeV) & $M_\pi^\text{val} L$ & $t_\text{sep}/a$ & $P_z$ & $N_\text{cfg}$ & $N_\text{meas}$  \\
\hline
a12m310 & 0.12 & $24^3\times 64$ & 310 & 683 & 4.55 & $\{6,7,8,9\}$ &  $\{3\}\frac{2\pi}{L}$  & 958  & $\{22922,45984,45984,61312\}$   \\
a12m220L & 0.12 & $40^3\times 64$ & 217 & 687 & 5.5 & $\{6,7,8,9\}$ &  $\{4,5\}\frac{2\pi}{L}$ & 840  & $\{13440,26880,26880,53760\}$ \\
a06m310 & 0.06 & $48^3\times 96$ & 319 & 690 & 4.52 & $\{12,14,16,18\}$ &  $\{3\}\frac{2\pi}{L}$    & 725  & $\{11600,23200,23200,46400\}$ \\
\end{tabular}
\end{ruledtabular}
\end{center}
\caption{\label{tab:params}
Ensemble information and parameters used in this calculation. $N_\text{meas}$ corresponds to the total number of measurements of the three-point correlators for $t_\text{sep}=\{0.72,0.84,0.96,1.08\}$~fm, respectively.
}
\end{table*}

In this work we use clover valence fermions with $N_f=2$+1+1 (degenerate up/down, strange and charm) highly improved staggered dynamical quarks (HISQ)~\cite{Follana:2006rc} in the sea, on ensembles generated by MILC Collaboration~\cite{Bazavov:2012xda}.
We use one step of hypercubic (HYP) smearing on the gauge links~\cite{Hasenfratz:2001hp} to suppress discretization effects, and the fermion-action parameters are tuned to recover the lowest pion mass of the staggered quarks in the sea.
Details can be found in Refs.~\cite{Rajan:2017lxk,Bhattacharya:2015wna,Bhattacharya:2015esa,Bhattacharya:2013ehc}.
We note that no exceptional configurations have been found among all the ensembles we use in this work~\cite{Rajan:2017lxk,Bhattacharya:2015wna,Bhattacharya:2015esa,Bhattacharya:2013ehc}.
The multigrid algorithm~\cite{Babich:2010qb,Osborn:2010mb} in the Chroma software package~\cite{Edwards:2004sx} is used to speed up the clover fermion inversion of the quark propagators.
We use Gaussian momentum smearing~\cite{Bali:2016lva} for both the light- and strange-quark fields $\psi(x) + \alpha \sum_j U_j(x) e^{ik\hat{e}_j} \psi(x+\hat{e}_j)$, where $k$ is the input momentum-smearing parameter, $U_j(x)$ are the gauge links in the $j$ direction, and $\alpha$ is a tunable parameter as in traditional Gaussian smearing.
Table~\ref{tab:params} summarizes the momenta, source-sink separations, and statistics used in this work.

On the lattice, we calculate both two-point and three-point quasi-PDF correlators: 
\begin{align}
&C_\text{2pt}(t_\text{sep}, \vec{P}) = \langle 0 |\int d^3\vec{y}\, e^{i \vec{y} \cdot \vec{P}} \overline{M}_\text{ps}(\vec{y},t_\text{sep}) M_\text{ps}(\vec{0},0) | 0 \rangle,\nonumber\\
&C_\text{3pt}(z, t,t_\text{sep},\vec{P}) =
\langle 0 |\int d^3y\, e^{i \vec{y} \cdot \vec{P}} \overline{M}_\text{ps}(\vec{y},t_\text{sep}) \nonumber\\
 &\hspace{3em} \times\bar{q}(z\hat{z}, t)
  \Gamma \prod_{x=0}^{z-1} U_z(x\hat{z}, t) q(\vec{0}, t)  
  M_\text{ps}(\vec{0},0) | 0 \rangle, 
\label{eq:kaon_2pt3pt}
\end{align}
where $C_\text{3pt}$ is the three-point correlator with $q=\{l,s\}$ quarks, $C_\text{2pt}$ is the two-point correlator, 
$M_\text{ps}= \bar{q}_1 \gamma_5 q_2$ is the pseudoscalar meson operator with $q_{1,2}$ being either the light- or strange-quark operator, 
$z$ is the length of the Wilson line, 
$U_\mu(\vec{x}, t)$ is a lattice gauge link.
As mentioned before, we choose Dirac spinor matrices $\Gamma=\gamma_t$ here as suggested in Refs.~\cite{Constantinou:2017sej,Chen:2017mie} to avoid mixing with the scalar matrix element. 
$t$ and $t_\text{sep}$ are the operator-insertion time and source-sink separation.
We choose the meson boost momentum $\vec{P}$ to lie along the $z$ direction and denote its magnitude as $P_z$.
All the source locations are randomly selected for each configuration; we shift to $t=0$ for convenience before the measurements are averaged.  

The matrix elements for the meson quasi-PDF are then extracted using multiple source-sink separations $t_\text{sep}$, removing excited-state contamination by performing ``two-simRR'' fits~\cite{Bhattacharya:2013ehc}:
\begin{align}
\label{eq:c3ptfitform}
C^\text{3pt}_{\Gamma}(P_z,t,t_\text{sep}) &=
   |{\cal A}_0|^2 \langle 0 | \mathcal{O}_\Gamma | 0 \rangle  e^{-E_0t_\text{sep}} \nonumber\\
   &+|{\cal A}_1{\cal A}_0| \langle 1 | \mathcal{O}_\Gamma | 0 \rangle  e^{-E_1 (t_\text{sep}-t)} e^{-E_0 t} \nonumber\\
   &+|{\cal A}_0{\cal A}_1| \langle 0 | \mathcal{O}_\Gamma | 1 \rangle  e^{-E_0 (t_\text{sep}-t)} e^{-E_1 t} \nonumber\\ 
   &+|{\cal A}_1|^2 \langle 1 | \mathcal{O}_\Gamma | 1 \rangle  e^{-E_1t_\text{sep}} + \ldots
\end{align}
at each meson boost momentum. 
The $E_0$ ($E_1$) and ${\cal A}_0$ (${\cal A}_1$)  are the ground- (excited-) state meson energy and overlap factors, extracted from the two-point correlators by fitting them to the form
\begin{equation}
C^\text{2pt}(P_z,t_\text{sep}) =
   |{\cal A}_0|^2 e^{-E_0 t_\text{sep}} + |{\cal A}_1|^2 e^{-E_1 t_\text{sep}}+\dots 
\end{equation}

\begin{figure}[htbp]
\includegraphics[width=.48\textwidth]{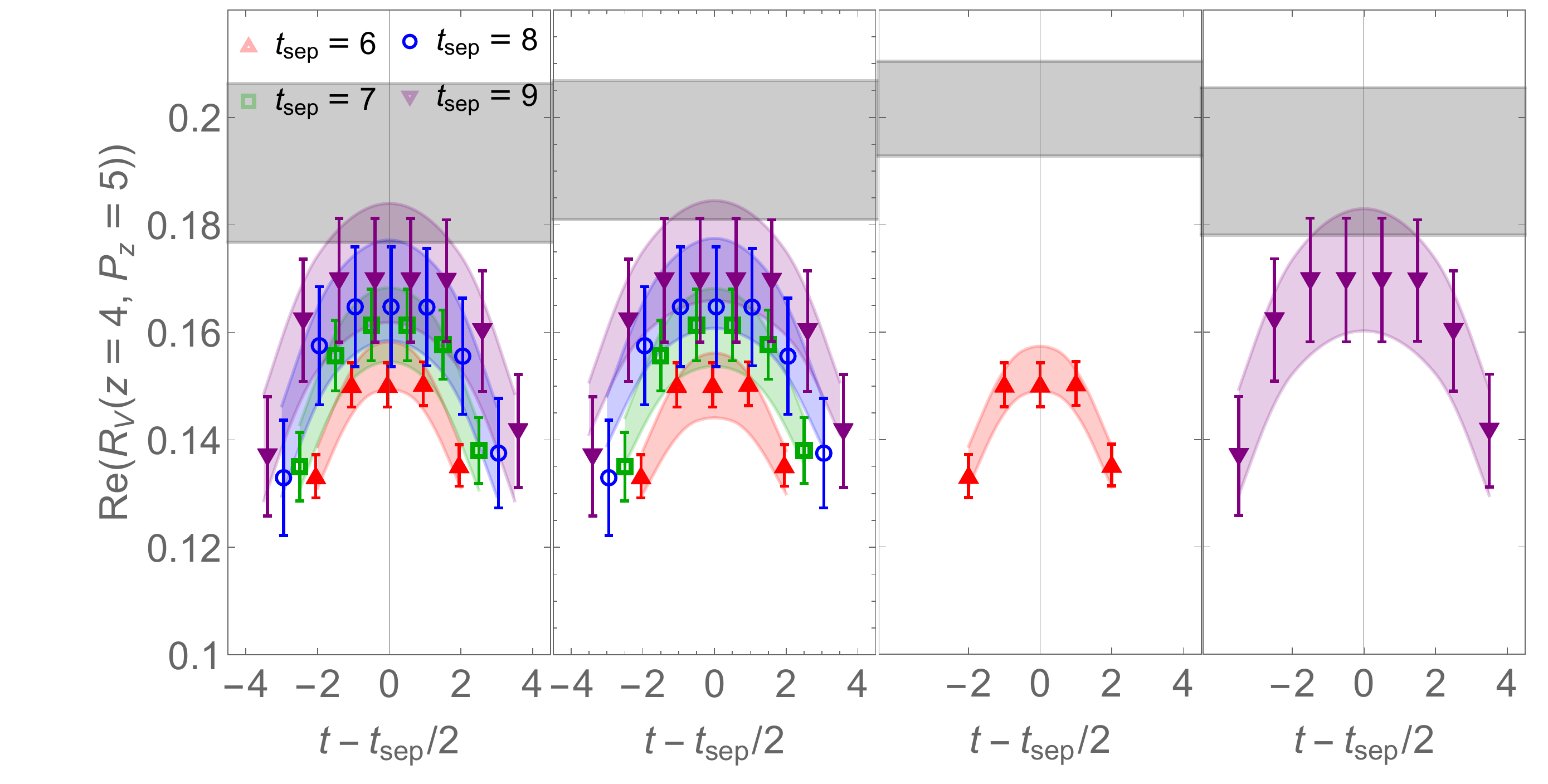} 
\caption{\label{fig:ExampleRatioPlots}
Example plots of the ratio of three- to two-point correlators as functions of the insertion time $t$ from the a12m220L ensemble. 
The real parts of the matrix elements are shown for kaon momentum $P_z \approx 1.7$~GeV and length of the Wilson line $z/a=4$, with curved bands showing fits using different source-sink separations $t_\text{sep}/a \in\{6,7,8,9\}$.
The gray bands indicate the final extracted ground-state matrix elements.
The left two plots show the fitted ratios $R$ and their corresponding ground-state matrix elements obtained via the ``two-simRR'' and ``two-sim''  methods, while the other two are ``two-state'' fits using only one $t_\text{sep}$.
The ground-state extractions are consistent across different choices of source-sink three-point inputs, as well as across different choices of analysis method.
}
\end{figure}

A few selected fits and the corresponding three-point ratio 
\begin{equation}
R_V(t_\text{sep},t) = \frac{C_\text{3pt}(t_\text{sep},t)}{ C_\text{2pt}(t_\text{sep}) }
   \end{equation}
are plotted from a subset of data on all three ensembles with $P_z = 5 \times 2\pi/L$ from the a12m220L ensemble in Fig.~\ref{fig:ExampleRatioPlots}; these use different $t_\text{sep}$, with source-sink separations from 0.72~fm to 1.08~fm. 
The leftmost plot shows  a ``two-simRR'' fit,  where all the $t_\text{sep}$ are fit simultaneously to all terms listed in Eq.~\ref{eq:c3ptfitform}; the plot to its right is a ``two-sim'' fit without the $\langle 1 | \mathcal{O}_\Gamma | 1 \rangle$ term. The extracted ground-state matrix elements are consistent between these two analysis methods. 
We also examine the fitted ground-state matrix elements from a two-state fit to selected $t_\text{sep}$ in the right two plots. The extracted ground-state matrix elements are also consistent among different $t_\text{sep}$, and agree with the simultaneous fits using all $t_\text{sep}$. 
The signal-to-noise ratio deteriorates significantly as $t_\text{sep}$ is increased, even though we have increased the number of measurements at larger source-sink separations. 
One can clearly see that the simultaneous fits well describe data from all $t_\text{sep}$, and the errors in the final extracted ground-state matrix-element are not over-constrained by the smallest $t_\text{sep}$ data. 
For the remainder of the paper, we only use the ``two-sim'' fits to obtain ground-state matrix elements for further processing.

\begin{figure}[htbp]
\includegraphics[width=.45\textwidth]{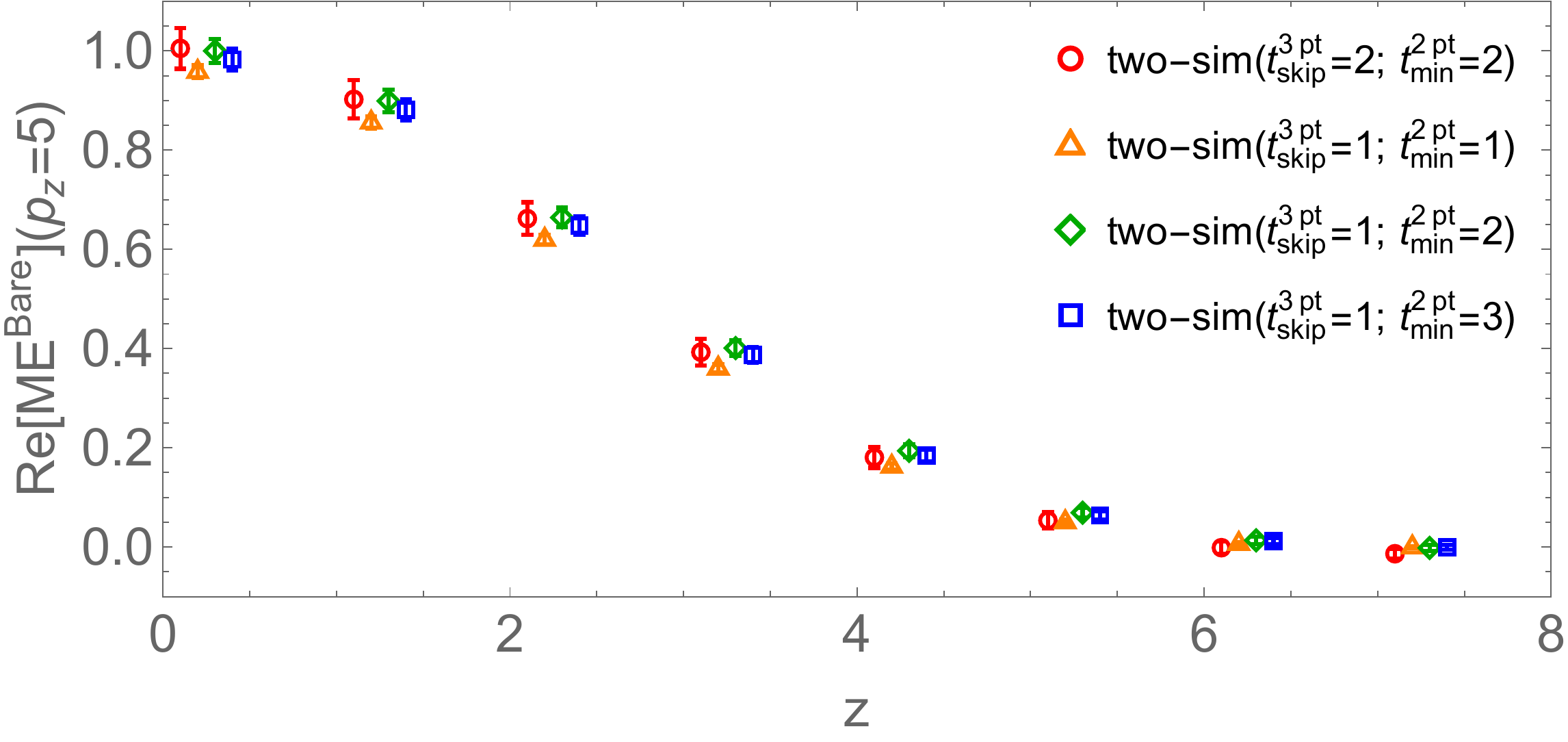}
\caption{\label{fig:FitCompPz5}
Comparison of the fitted kaon ground-state matrix elements as functions of Wilson-line length $z$ (in lattice units) from the a12m220L ensemble with $P_z\approx 5 \times \frac{2\pi}{L}$ and pion mass of 220~MeV using ``two-sim'' fits and varying the fit range of the two-point ($t_\text{min}^\text{2pt}$ corresponding to fit range of $[t_\text{min}^\text{2pt}, N_t/4]$ )  and three-point correlators ($t_\text{skip}^\text{3pt}$ corresponding to fit range of $[t_\text{skip}^\text{3pt}, t_\text{sep}-t_\text{skip}^\text{3pt}]$).
}
\end{figure}

To make sure that our extracted ground-state matrix elements are insensitive to the fit range used in correlators, we vary the fit range used for the two- and three-point correlators and compare the extracted matrix elements.
Figure~\ref{fig:FitCompPz5} shows an example result from one of the ensembles, where we can see that the extracted ground-state matrix elements are stable across different fit-range choices among two-point and three-point correlators.

\begin{figure}[htbp]
\includegraphics[width=.45\textwidth]{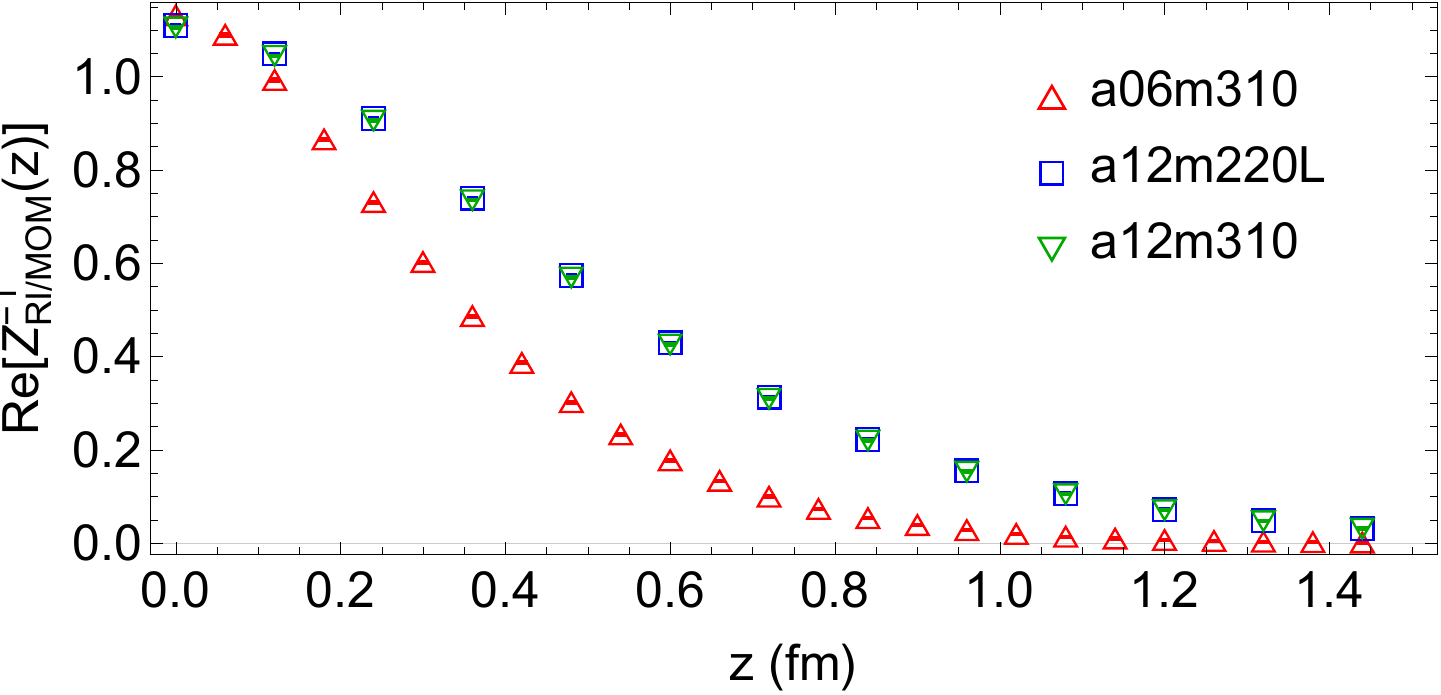}
\caption{The inverse renormalization factor from all three ensembles as functions of Wilson-line displacement $z$ with RI/MOM renormalization scales $\mu_R = 2.4$~GeV and $p_z^R=0$.} \label{fig:NPR-plots}
\end{figure}

Once we obtain the bare ground-state matrix elements,
\begin{equation}
h_{\tilde n_t}(\lambda\tilde n)=\frac{P_z}{P_t}\langle K(P)\left|\bar\psi_q(\lambda {\tilde n})\slashed {n_t} W(\lambda {\tilde n},0)\psi_q(0)\right|K(P)\rangle,
\end{equation}
the next step is to renormalize them as 
\begin{equation}\label{hR}
 h_{\tilde n_t, R}(\lambda\tilde n) = Z^{-1}(p_z^R, 1/a, \mu_R) h_{\tilde n_t}(\lambda\tilde n),
\end{equation}
with the RI/MOM renormalization factor being defined as
\begin{align}\label{hRx}
&Z(p^R_z, 1/a, \mu_R) =\nonumber\\
&\left.\frac{\Tr[\slashed p \sum_s \langle p,s| \bar\psi_f(\lambda \tilde n) \slashed{\tilde n_t} W(\lambda\tilde n,0) \psi_f(0)|p,s\rangle]}
{\Tr[\slashed p  \sum_s \langle p,s| \bar\psi_f(\lambda \tilde n) \slashed{\tilde n_t} W(\lambda\tilde n,0) \psi_f(0) |p,s\rangle_\text{tree}]} \right|_{\tiny\begin{matrix}p^2=-\mu_R^2 \\ \!\!\!\!p_z=p^R_z\end{matrix}}.
\end{align} 
In Fig.~\ref{fig:NPR-plots} we show the RI/MOM renormalization factors calculated from all three ensembles. 
As can be seen from the figure, the dependence of the renormalization factors on lattice spacing is significant, because they serve as counterterms to cancel the UV divergence of the bare matrix elements; contrariwise, the dependence on pion mass is negligible.

\section{Results and Discussion}\label{sec:Results}
\begin{figure}[htbp]
\includegraphics[width=.4\textwidth]{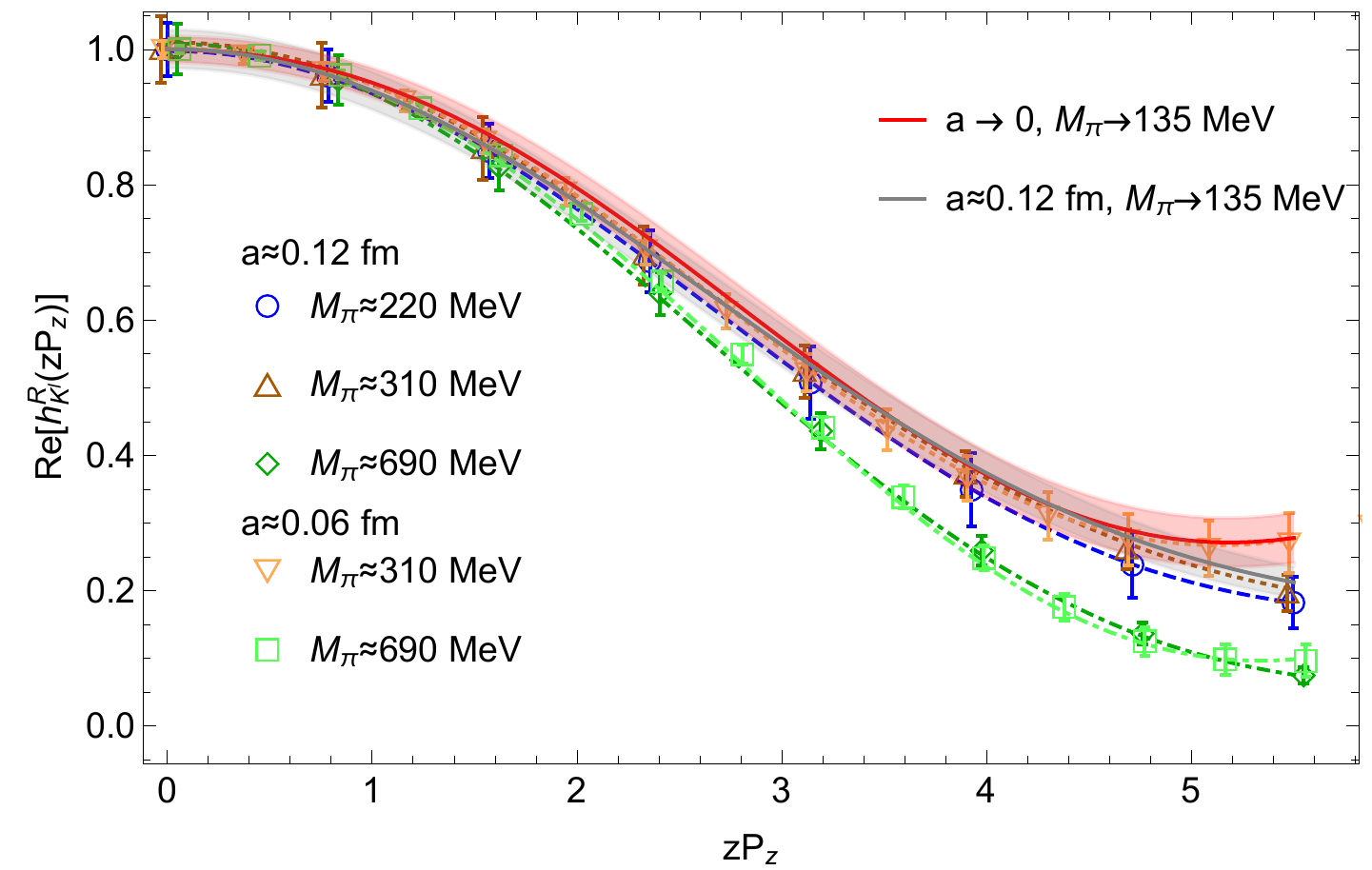}
\includegraphics[width=.4\textwidth]{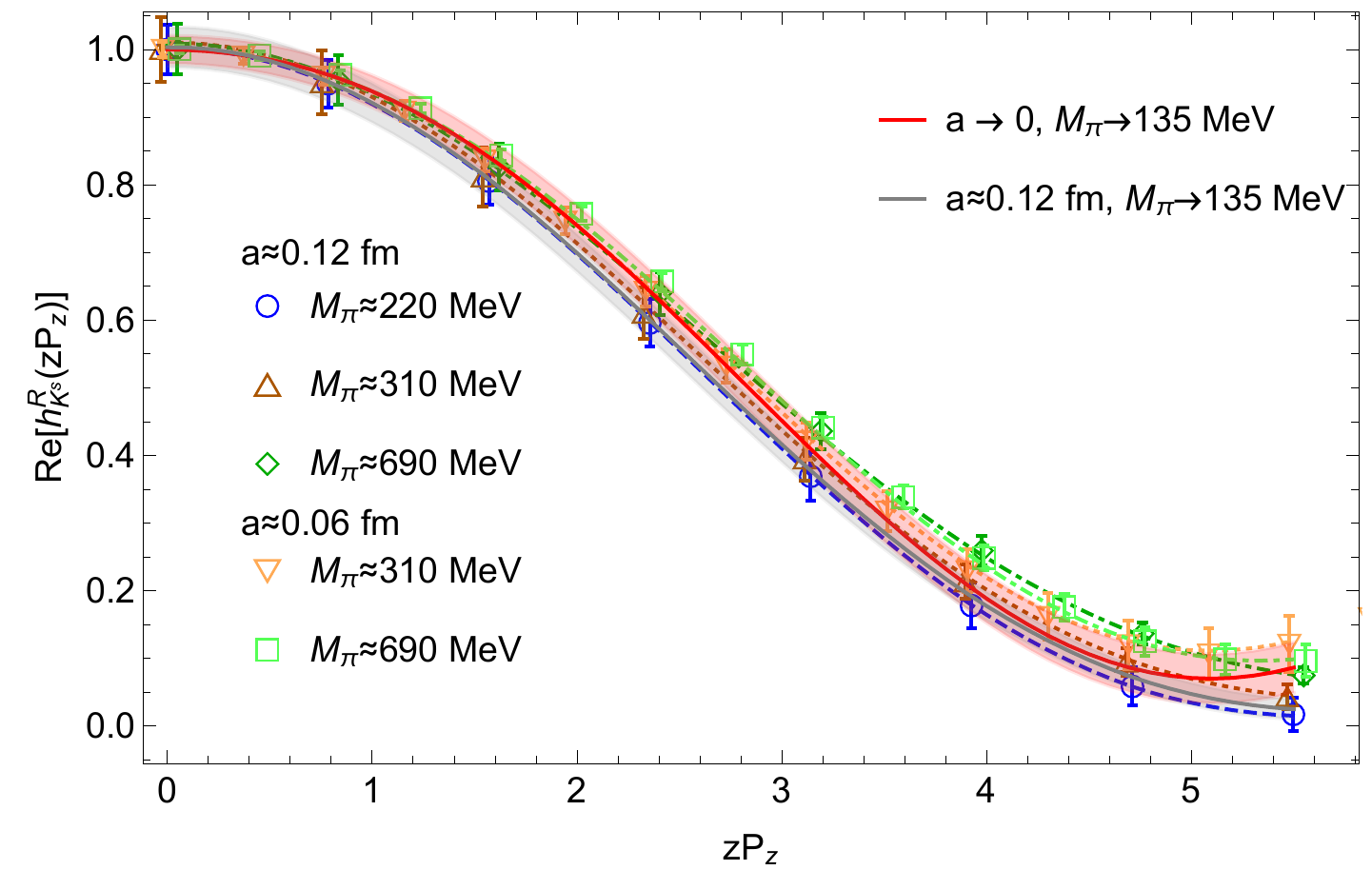}
\caption{The renormalized matrix elements of the light (top) and strange (bottom) valence-quark contributions to the kaon PDFs as functions of dimensionless $zP_z$ for all three ensembles, and the chiral  (gray band) and chiral-continuum (pink band) extrapolation. 
Both plots show fixed $P_z=1.3$~GeV. 
}
\label{fig:PhysME}
\end{figure}

In this work, we compute the RI/MOM renormalization factors at $\mu^R=\sqrt{-p^2}=2.4$~GeV and $p_z^R =p_z= 0$ with $p$ denoting the off-shell quark momentum. For $p_z^R = 0$, the renormalization factors are real. 
Figure~\ref{fig:PhysME} shows the renormalized matrix elements for the light valence quark of the kaon. We observe a small pion-mass dependence for the two $a\approx 0.12$~fm ensembles between the ensembles with 220- and 310-MeV pions,
and a benign lattice-spacing dependence between $a\approx 0.06$ and 0.12~fm in most regions of $zP_z$. 
Next, we perform a chiral-continuum extrapolation to obtain the renormalized matrix elements at physical pion mass. 
We use a simple ansatz to combine our data from 220, 310 and 690~MeV: $h^R_{i}(P_z, z, M_\pi) = c_{0,i} + c_{1,i} M_\pi^2 + c_{a} a^2$ with $i=K, \pi$. 
Mixed actions, with light and strange quark masses tuned to reproduce the lightest sea light and strange pseudoscalar meson masses, can suffer from additional systematics at $O(a^2)$~\cite{Chen:2007ug}; such artifacts are accounted for by the $c_{a}$ coefficient. We find all the $c_{a}$ to be consistent with zero. 
Example plots of the chiral ($c_a=0$ with only $a\approx 0.12$~fm data) and chiral-continuum extrapolations of the light-valence kaon can be found in Fig.~\ref{fig:PhysME}, where the results from individual ensembles are shown as lines, while the extrapolated results at physical pion mass are shown as pink (chiral) and gray (chiral-continuum) bands. Overall, the extrapolated matrix elements are consistent with the 310- and 220-MeV results, but can be significantly different from the 690-MeV ones due to the heavy mass.

With the matrix elements at physical pion-mass, we can extract the pion and kaon PDFs through Eq.~\ref{matching} using a parametrization-and-fit procedure, as used in Ref.~\cite{Izubuchi:2019lyk}. We take the commonly used analytical form
\begin{align}
f_{m,n,c} (x) =
  \frac{x^m (1-x)^n (1+c\sqrt{x})}
  {B(m+1,n+1) + c B\left(m+\frac{3}{2},n+1\right)}
\label{eq:PDFfitFunc}
\end{align}
where $B(m+1,n+1)=\int_0^1 dx\,x^m (1-x)^n$ is the beta function, which normalizes the distribution such that the area under the curve is unity. 
We study the uncertainty by comparing the PDF results between the two-parameter fit ($c=0$) and the form with the additional $\sqrt{x}$ term.  
By applying the matching~\cite{Chen:2018xof,Stewart:2017tvs} to the parametrized $\overline{\text{MS}}$ PDF at 2~GeV with the meson-mass correction from Ref.~\cite{Chen:2016utp} included, we are able to determine the unknown parameters $m, n$ from the RI/MOM renormalized quasi-PDF. 

\begin{figure}[htbp]
\includegraphics[width=.45\textwidth]{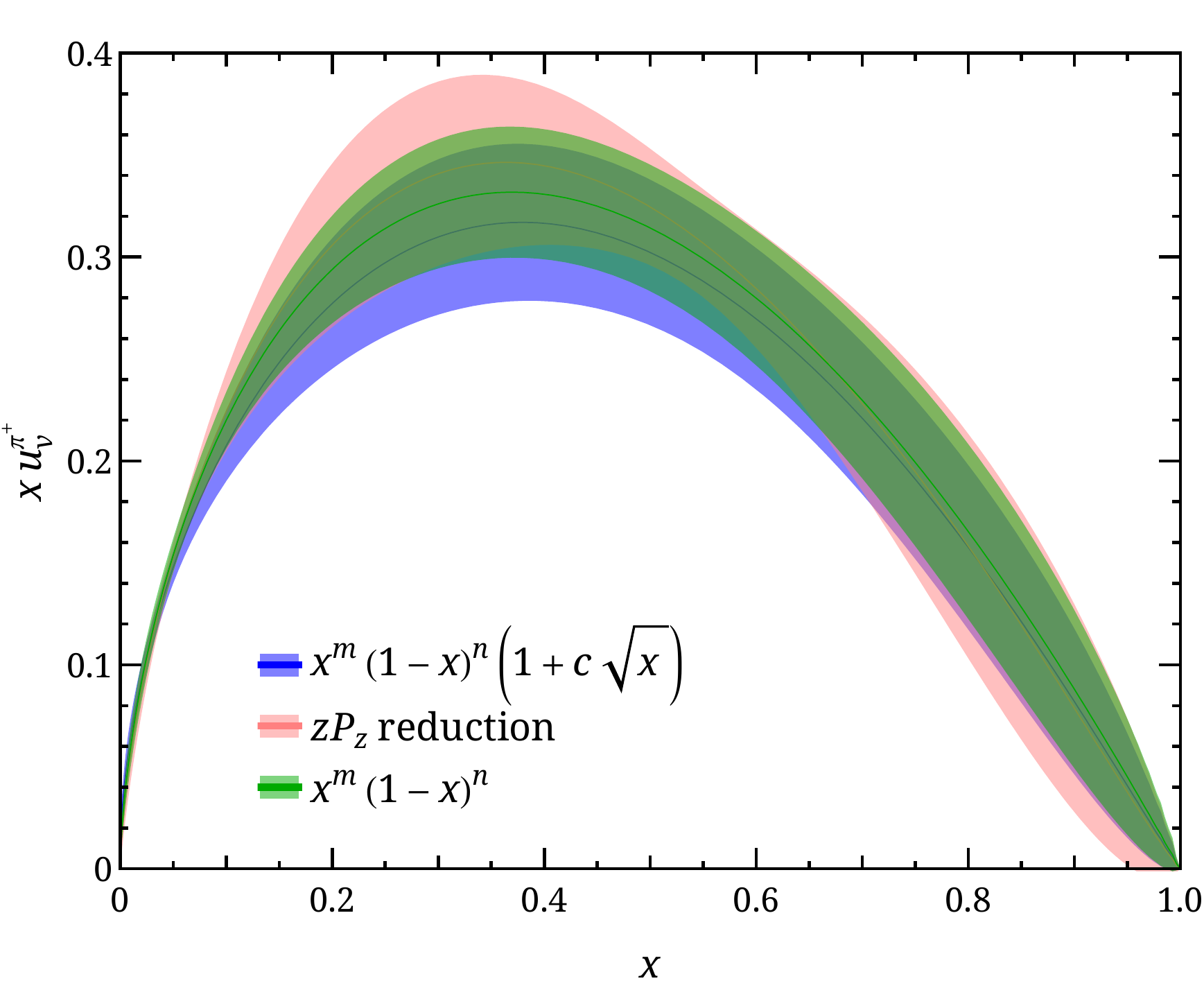}
\includegraphics[width=.45\textwidth]{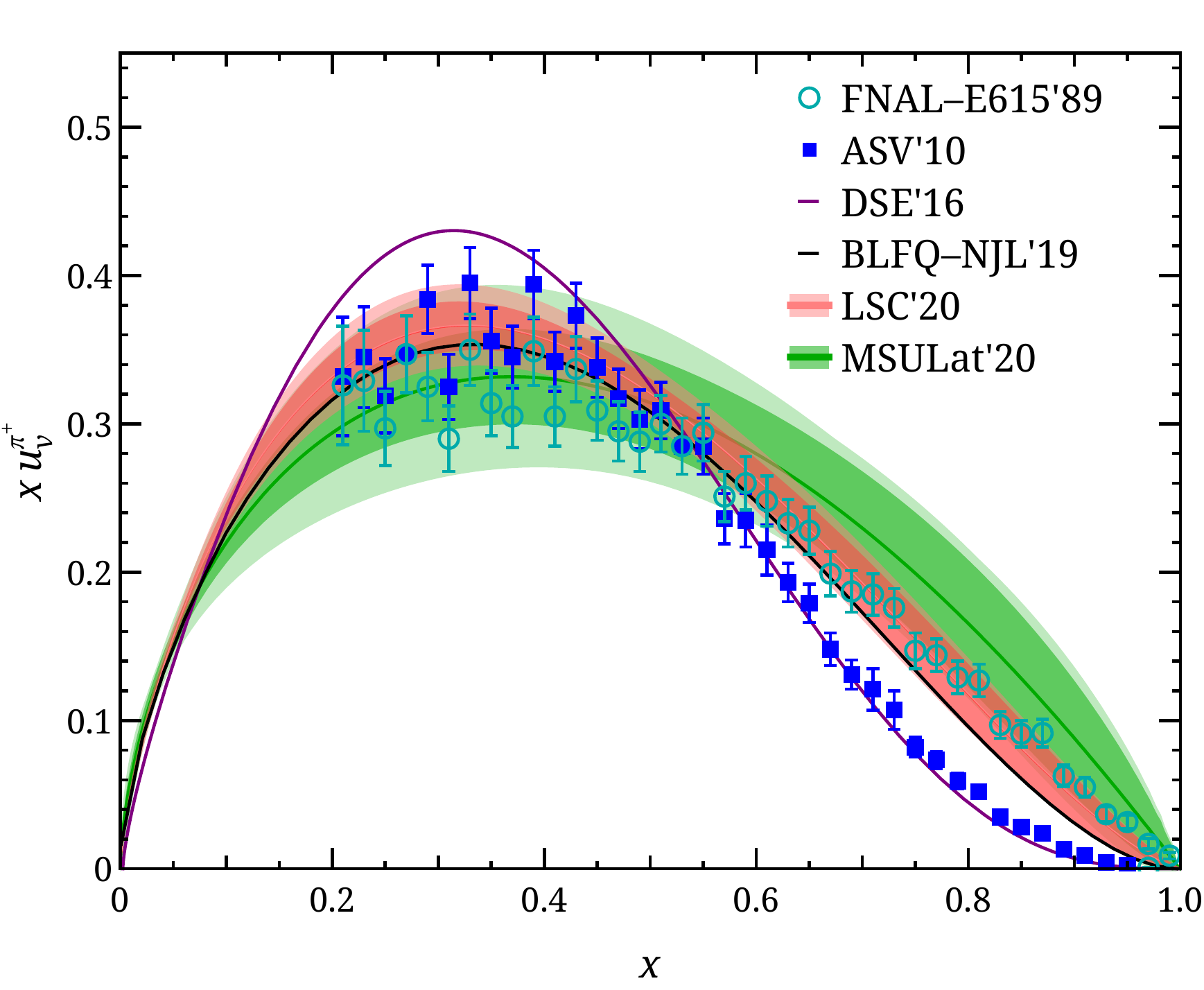}
\caption{
(Top) Comparison of our valence-quark distribution of the pion (top)  at a scale of $27\text{ GeV}^2$ using the two-parameter form ($c = 0$) 
of Eq.~\ref{eq:PDFfitFunc} with the full range of $zP_z$ data (green), a $1/8$-reduction of $zP_z$ data (pink) and Eq.~\ref{eq:PDFfitFunc} with the full range of $zP_z$ data (blue); the difference among different data choices and fit form are smaller than the statistical errors. 
(Bottom) Comparison of our result (labeled ``MSULat'20'', shown as a green band) with analysis from experimental data and calculations from other methods (see the text for details) for $x u_v^\pi$ as a function of $x$ at scale of
$27\text{ GeV}^2$.
}
\label{fig:xpion}
\end{figure}

The top part of Fig.~\ref{fig:xpion} shows the valence-quark distribution of the pion obtained using two- (green band) and three-parameter fits (blue band).
We also study the dependence on the maximum available Wilson-line displacement;
we reduce the maximum displacement by one-eighth and use the two-parameter fit.
The result is shown as a pink band on the same plot.
In both studies, we obtain a slightly wider band, as anticipated, due to the reduced number of degrees of freedom; 
overall, the shift of the central values of the distribution is small compared with the statistical error.
In the rest of this work, we will take the two-parameter fit with full set of data as main result, and take the maximal difference in the central values from the other two fits as a the size of the systematic uncertainty.

Our leading moments from the pion distribution are $\langle x \rangle_v = 0.281(23)_\text{stat}(14)_\text{syst}$, $\langle x^2 \rangle_v = 0.142(18)_\text{stat}(6)_\text{syst}$, $\langle x^3 \rangle_v =0.086(15)_\text{stat}(4)_\text{syst}$, which are consistent with the traditional moment approach done by ETMC using $N_f=2+1+1$ twisted-mass fermions with pion masses in the range of 230 to 500~MeV, renormalized at 2~GeV; see Table V in Ref.~\cite{Oehm:2018jvm} with $\langle x \rangle$ ranging 0.23--0.29 and $\langle x^2 \rangle$ ranging 0.11--0.18. 
Figure~\ref{fig:xpion} shows our final results for the pion valence distribution at physical pion mass ($u_v^{\pi^+}$) multiplied by Bjorken-$x$ as a function of $x$. We evolve our results to a scale of 27$\text{ GeV}^2$ using the 
NNLO DGLAP equations from the Higher-Order Perturbative Parton Evolution Toolkit (HOPPET)~\cite{Salam:2008qg} to compare with other results. 
Our result approaches large-$x$ as $(1-x)^{1.01}$ and is consistent with the
original analysis of the FNAL-E615 experiment data~\cite{Conway:1989fs}, whereas there is tension with the $x>0.6$ distribution from the re-analysis of the FNAL-E615 experiment data using 
next-to-leading-logarithmic threshold resummation effects in the calculation of the Drell-Yan cross section~\cite{Aicher:2010cb} (labeled as ``ASV'10''), which agrees better with the distribution from Dyson-Schwinger equations (DSE)~\cite{Chen:2016sno}; both prefer the form $(1-x)^2$ as $x \to 1$. 
An independent lattice study of the pion valence-quark distribution~\cite{Sufian:2020vzb}, also extrapolated to physical pion mass, using the ``lattice cross sections'' (LCSs)~\cite{Ma:2014jla}, reported similar results to ours. 
Our lowest 3 moments at the scale of $27\text{ GeV}^2$ are $0.225(18)_\text{stat}(10)_\text{syst}$, $0.100(13)_\text{stat}(5)_\text{syst}$, $0.056(10)_\text{stat}(2)_\text{syst}$,
which are consistent with the moments (0.23, 0.094, 0.048) from chiral constituent quark model~\cite{Watanabe:2017pvl}.

\begin{figure}[tbp]
\includegraphics[width=.42\textwidth]{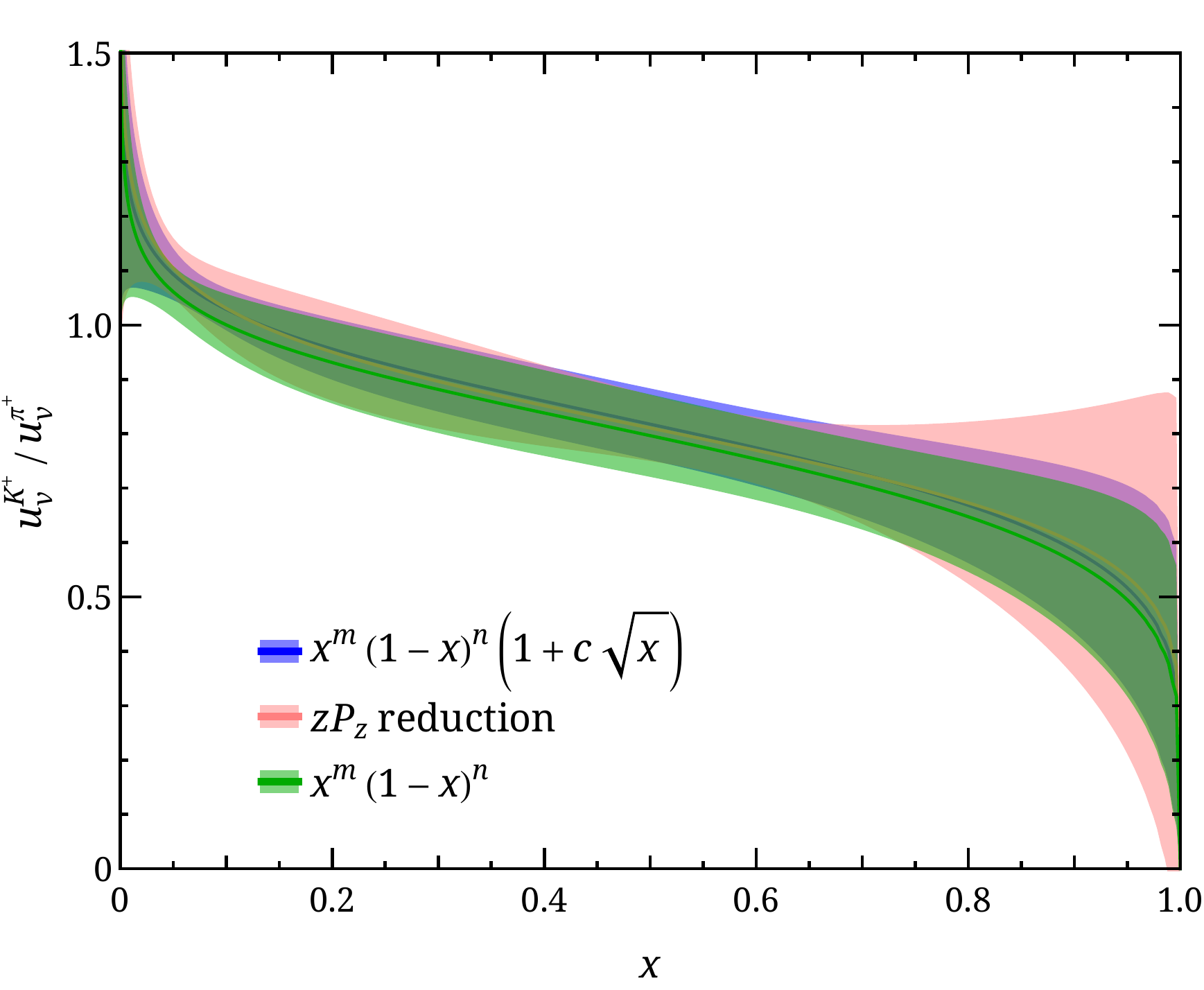}
\includegraphics[width=.42\textwidth]{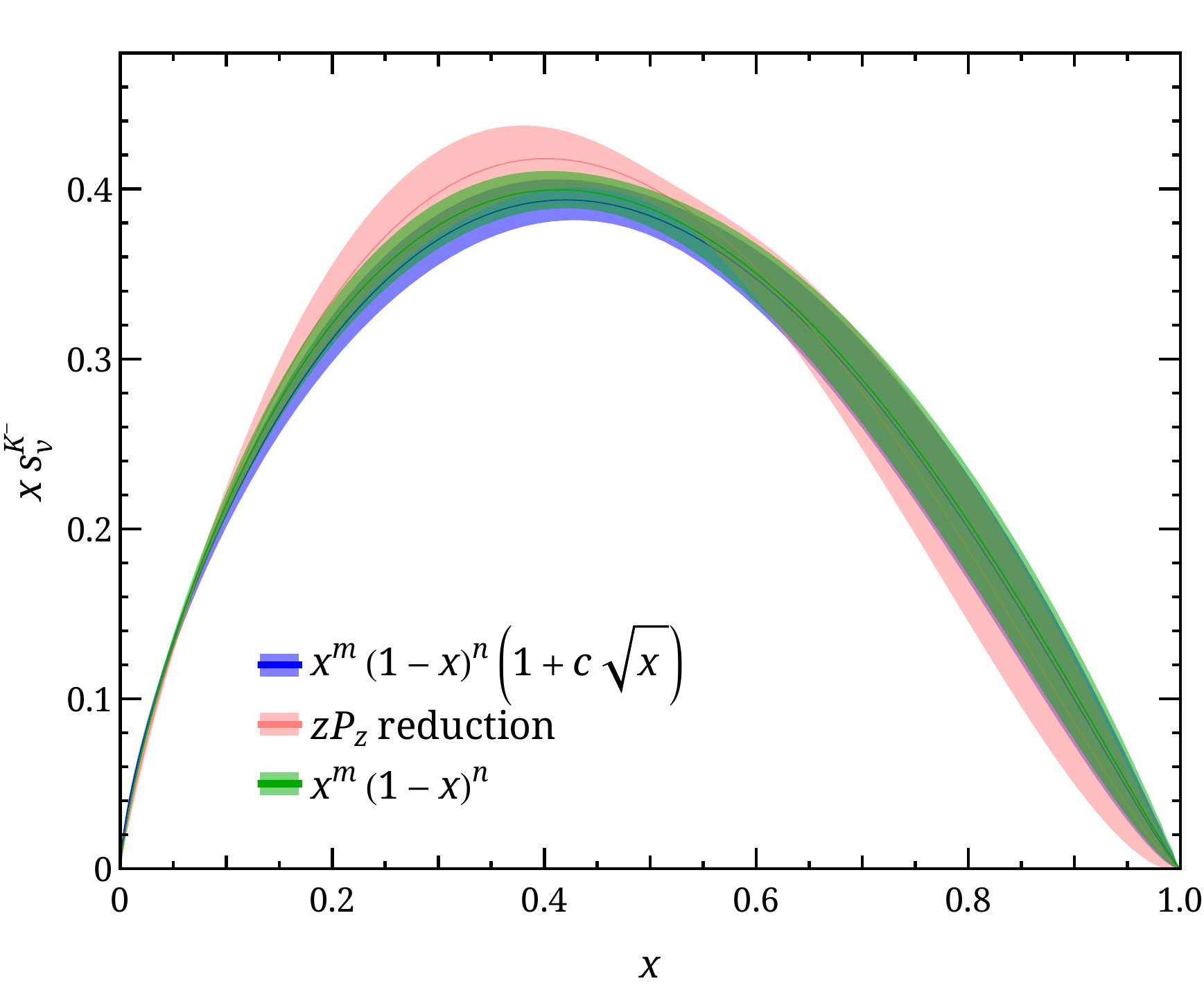}
\caption{Comparison of our main results on the ratio of the light-quark valence distribution of the kaon to that of the pion (top) and $x \overline{s}_v^K(x)$ as a function of $x$ (bottom) at a scale of $27\text{ GeV}^2$ using the two-parameter form  ($c=0$) of Eq.~\ref{eq:PDFfitFunc} with the full range of $zP_z$ data (green), a $1/8$-reduction of $zP_z$ data (pink) and three-parameter fit with the full range of $zP_z$ data (blue); the differences among different data choices and fit form are smaller than the statistical errors. 
} \label{fig:PDFs-various}
\end{figure}

\begin{figure}[tbp]
\includegraphics[width=.45\textwidth]{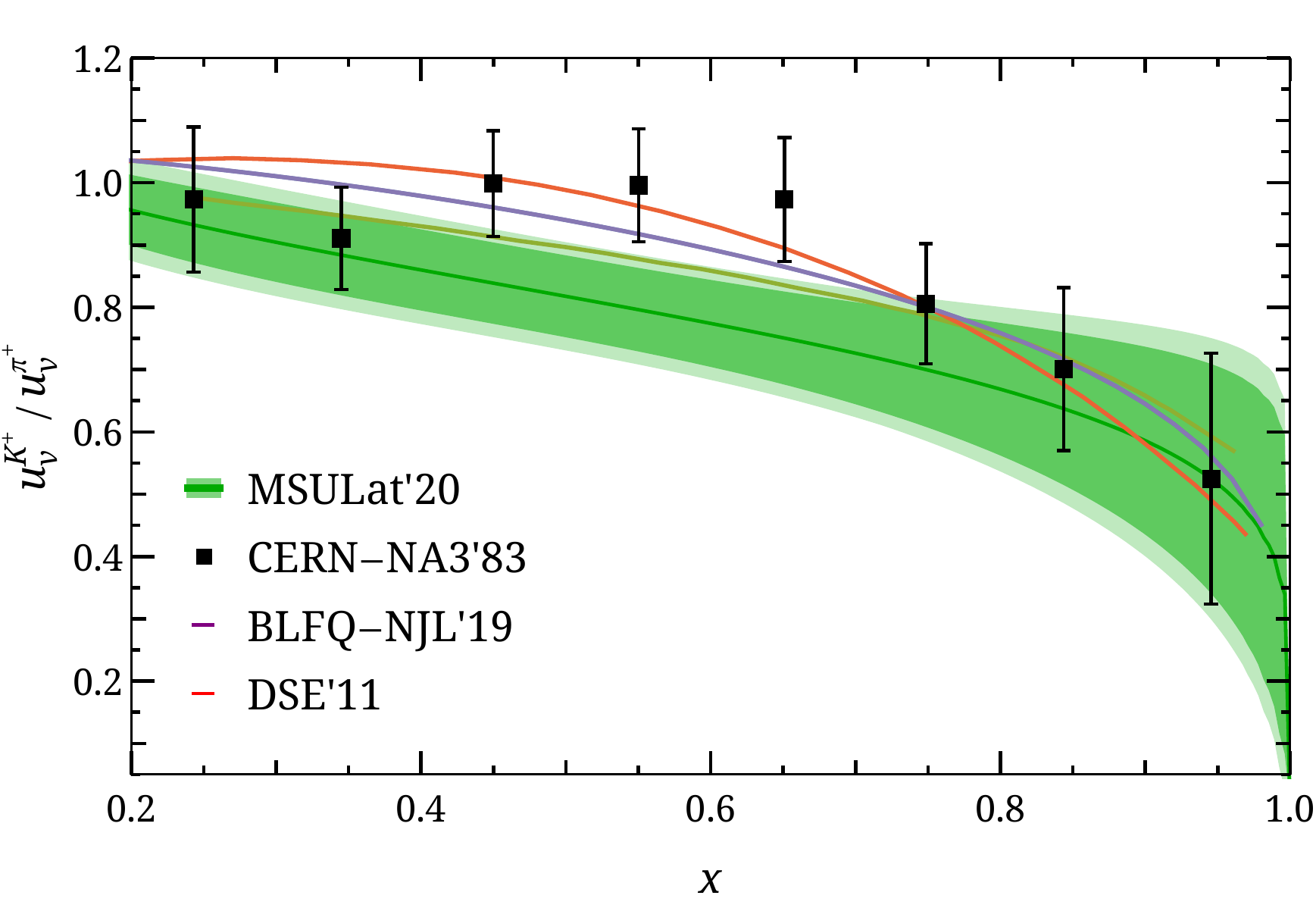}
\includegraphics[width=.45\textwidth]{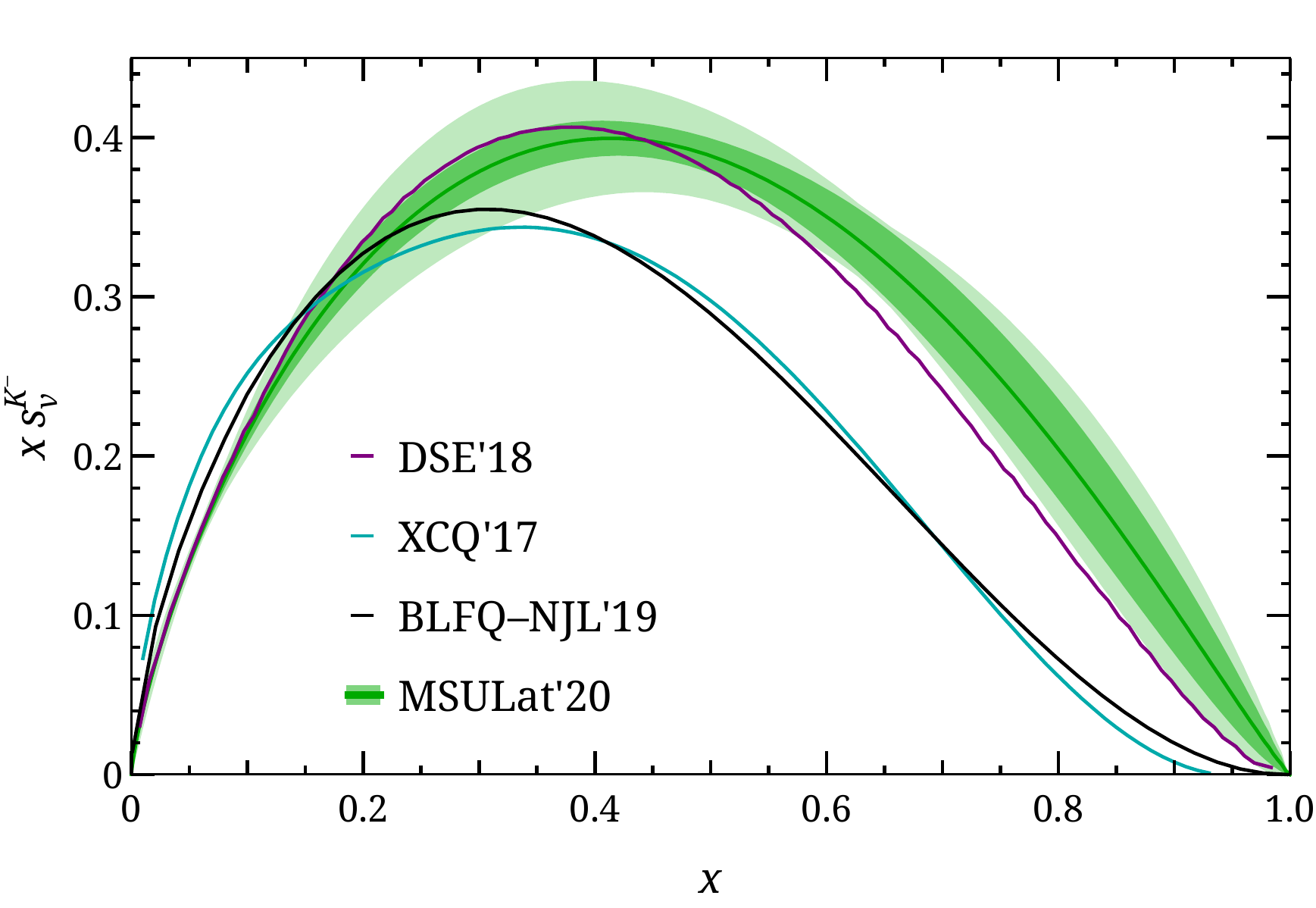}
\caption{Results on the ratio of the light-quark valence distribution of kaon to that of pion (top) and for $x \overline{s}_v^K(x)$ as a function of $x$ (bottom) at scale of $27\text{ GeV}^2$, both labeled ``MSULat'20'', along with results from relevant experiment/other calculations (see the text for details).
}
\label{fig:ratios}
\end{figure}

Figure~\ref{fig:PDFs-various} shows comparison plots to examine the impact of the fit form (shown as green and blue bands) on the ratio of the light-quark valence distribution of kaon to that of the pion and on the antistrange valence distribution of kaon; we find that the difference is small. 
We further compare the same results using the two-parameter fit form of Eq.~\ref{eq:PDFfitFunc} but with data truncated from the max $zP_z$ by one eighth, shown as pink bands in Fig.~\ref{fig:PDFs-various}. 
We added the difference as a systematic uncertainty in Figs.~\ref{fig:xpion} and \ref{fig:ratios}. 

Figure~\ref{fig:ratios} shows the ratios of the light-quark distribution in the kaon to the one in the pion ($u_v^{K^+}/u_v^{\pi^+}$). 
When comparing our result with the original experimental determination of the valence quark distribution via the Drell-Yan process by 
NA3 Collaboration~\cite{Badier:1983mj} in 1982, we found good agreement between our results and the data. Our result approaches $0.4$ as $x \to 1$ and agrees nicely with other analyses, such as constituent quark model~\cite{Gluck:1997ww}, 
the DSE approach (``DSE'11'')~\cite{Nguyen:2011jy},
and basis light-front quantization with color-singlet Nambu--Jona-Lasinio interactions (``BLFQ-NJL'19'')~\cite{Lan:2019vui}. 
Our lowest 3 moments for $u_v^{K^+}$ are $0.192(8)_\text{stat}(6)_\text{syst}$, $0.080(7)_\text{stat}(6)_\text{syst}$, $0.041(6)_\text{stat}(4)_\text{syst}$, respectively, which are within the discrepancies of various QCD model estimates of
0.23, 0.091, 0.045 from chiral constituent-quark model~\cite{Watanabe:2017pvl}
and 
0.28, 0.11, 0.048 from DSE~\cite{Chen:2016sno}. 
Our prediction for $x s_v^{K}$ is also shown in Fig.~\ref{fig:ratios} with the lowest 3 moments of $s_v^{K}$ being $0.261(8)_\text{stat}(8)_\text{syst}$, $0.120(7)_\text{stat}(9)_\text{syst}$, $0.069(6)_\text{stat}(8)_\text{syst}$, respectively; the moment results are within the ranges of the QCD model estimates
from chiral constituent-quark model~\cite{Watanabe:2017pvl}
(0.24, 0.096, 0.049)
and 
DSE~\cite{Chen:2016sno} (0.36, 0.17, 0.092).

\section{Conclusion}\label{sec:Conclusion} 
In this work, we presented the first direct lattice-QCD calculation of the $x$ dependence of the kaon parton distribution functions using two lattice spacings, multiple pion masses ($M_{\pi,\text{min}} = 217$~MeV) and $M_\pi L \in \{4.5, 5.5\}$ with high statistics, $N_\text{meas} \in \{11,61\}$ thousands and $N_\text{cfg} \in \{725,958\}$. 
Our valence-quark pion distribution is in good agreement with the one obtained by JLab/W\&M group using LSC methods and extrapolated to the physical pion mass. 
The ratios of the light-quark valence distribution in the kaon to the one in pion, $u_v^K/u_v^\pi$, were found to be consistent with the original CERN NA3 experiments. 
We also made predictions for the strange-quark valence distribution of the kaon, $s_v^K(x)$, determining that it is close to the DSE result~\cite{Bednar:2018mtf}.

\section*{Acknowledgments}
We thank the MILC Collaboration and RBC Collaboration for sharing the lattices used to perform this study. The LQCD calculations were performed using the Chroma software
suite~\cite{Edwards:2004sx}.
This research used resources of the National Energy Research Scientific Computing Center, a DOE Office of Science User Facility supported by the Office of Science of the U.S. Department of Energy under Contract No. DE-AC02-05CH11231 through ERCAP; 
facilities of the USQCD Collaboration, which are funded by the Office of Science of the U.S. Department of Energy, 
and supported in part by Michigan State University through computational resources provided by the Institute for Cyber-Enabled Research (iCER). 
ZF, HL and RZ are supported by the US National Science Foundation under grant PHY 1653405 ``CAREER: Constraining Parton Distribution Functions for New-Physics Searches''. JWC is partly supported by the Ministry of Science and Technology, Taiwan, under Grant No. 108-2112-M-002-003-MY3 and the Kenda Foundation. JHZ is supported in part by National Natural Science Foundation of China under Grant No. 11975051, and by the Fundamental Research Funds for the Central Universities.

\ifx\@bibitemShut\undefined\let\@bibitemShut\relax\fi
\makeatother


\end{document}